
\magnification=1200
\baselineskip=18pt
\def\ao{\~ao\ }

\def\[{\c c}
\def\\{\'\i }
\def\sqr#1#2{{\vcenter{\hrule height.#2pt
     \hbox{\vrule width.#2pt height#1pt \kern#1pt
      \vrule width.#2pt} \hrule height.#2pt}}}

\def\sq{{\sqcap\!\!\!\!\sqcup\,}}
\font\titlea=cmb10 scaled\magstep1
\font\titleb=cmb10 scaled\magstep2

\hfill IF-UFRJ-05/94
\bigskip
\centerline{\titleb The question of anomalies in the Chern-Simons theory}
\centerline{\titleb coupled to matter fields}
\vskip 2.0 cm
\centerline{R. Amorim$\,^\ast$ and J. Barcelos-Neto$\,^\dagger$}
\bigskip
\centerline{\it Instituto de F\\sica}
\centerline{\it Universidade Federal do Rio de Janeiro}
\centerline{\it RJ 21945-970 - Caixa Postal 68528 - Brazil}
\vskip 1.5 cm
\centerline{\titlea Abstract}
\bigskip
We study the Chern-Simons theory coupled to matter field by means of
an effective Lagrangian obtained from the Batalin-Fradkin-Vilkovisky
formalism. We show that there is no rotational anomaly for any proper
gauge we choose.

\vfill
\noindent PACS: 03.70.+k, 11.15.-q, 11.30.-j.
\bigskip
\hbox to 3.5 cm {\hrulefill}\par
\item{($\ast$)} Electronic mail: ift01001 @ ufrj
\item{($\dagger$)} Electronic mails: ift03001 @ ufrj and barcelos @
vms1.nce.ufrj.br

\eject

{\titlea I. Introduction}
\bigskip
There has been much controversy on the rule of gauge fixation in the
appearance of fractional spin and statistics in the context of
Chern-Simons (CS) gauge theory coupled to charged matter fields~[1].
Several authors have claimed that these theories exhibit angular
momentum anomaly for some specific gauge~[2]. In a recent work,
however, Banerjee has shown that this kind of anomaly does not
exist~[3] if an specific gauge choice is not made.  He used the Dirac
formalism~[4], without fixing the gauge and following an approach where
the energy momentum tensor is properly modified by introducing
first-class constraints by means of Lagrange multipliers~[5]. These
are fixed by imposing the closure of the Poincar\'e algebra under the
Dirac bracket spectrum. It might be opportune to emphasize that these
terms are introduced by hand and that there are no effective
Lagrangian that can both lead to it and still correctly describes the
initial theory.

\medskip
The purpose of the present paper is twofold. First we use Hamiltonian
path integral formalism due to Batalin, Fradkin and Vilkovisky
(BFV)~[6] to obtain an effective Lagrangian without choosing any
specific form for the gauge fixing function. After that, with this
generic Lagrangian and a careful analysis of the Noether currents and
their relation to momenta, we show that the closure of the Poincar\'e
algebra can be achieved without anomalies for any gauge condition.
The main point it that  when one changes the gauge-fixing choice,
momenta also change in order to preserve the symmetry.

\medskip
Our paper is organized as follows: In Sec. II we make a brief
analysis of the canonical procedure in order to obtain the
interesting structure of constraints of the theory. In Sec. III we
discuss the elimination of the second class constraints in the path
integral formulation of the BFV  formalism and the obtainment of the
effective Lagrangian. Sec. IV contains the discussion of the closure
of the Poincar\'e algebra and Sec. V is devoted to some comments and
conclusions. We have also included two appendices in order to better
clarify some points of the paper.

\vfill\eject
{\titlea II. Brief review of the canonical procedure}
\bigskip
Pure CS vectorial bosons coupled to complex scalar fields can be
described by the Lagrangian density~[7]

$${\cal L}=(D_\mu\phi)^\dagger(D^\mu\phi)
+{\kappa\over4\pi}\,\epsilon^{\mu\nu\rho}\,
A_\mu\partial_\nu A_\rho\,,\eqno(2.1)$$

\noindent where $D_\mu=\partial_\mu+iA_\mu$ is the covariant
derivative and  $\epsilon^{\mu\nu\rho}$ is the
completely antisimmetric symbol (with
$\epsilon^{012}=\epsilon_{012}=1$). $\phi$, $\phi^\dagger$ and $A_\mu$
represent respectively complex scalar fields and vectorial gauge
bosons.  We adopt the Minkowsky metric tensor in $2+1$ spacetime
dimensions as $\eta_{\mu\nu}={\rm diag.}\,(+1,-1,-1)$.

\medskip
This theory leads to the following set of (primary and secondary)
constraints~[4,~5]

$$\eqalignno{\chi^0&=\pi^0\,,&(2.2a)\cr
\chi^i&=\pi^i-{\kappa\over4\pi}\,
\epsilon^{ij}\,A_j\,,&(2.2b)\cr
\psi&=i(\pi\phi-\pi^\dagger\phi^\dagger)
+{\kappa\over2\pi}\,\epsilon^{ij}\partial_iA_j\,,
&(2.2c)\cr}$$

\bigskip
\noindent where

$$\eqalignno{\pi^\mu&={\kappa\over4\pi}\,
\epsilon^{0\mu\nu}\,A_\nu\,,&(2.3a)\cr
\pi&=(D_0\phi)^\dagger
=\dot\phi^\dagger-iA_0\phi^\dagger\,,&(2.3b)\cr
\pi^\dagger&=D_0\phi=\dot\phi+iA_0\phi&(2.3c)\cr}$$

\bigskip
\noindent are the canonical momenta conjugate to $A_\mu$, $\phi$ and
$\phi^\dagger$, respectively. Let us next write down the total
Hamiltonian density

$${\cal H}={\cal H}_c+\lambda_0\chi^0+\lambda_i\chi^i
+\lambda\psi\,,\eqno(2.4)$$

\bigskip
\noindent where ${\cal H}_c$ is the canonical Hamiltonian density

$${\cal H}_c=\pi^0\dot A_0+\pi^i\dot A_i+\pi\dot\phi
+\pi^\dagger\dot\phi^\dagger-{\cal L}\,.\eqno(2.5)$$

\bigskip
\noindent Velocities $\dot A_0$ and $\dot A_i$ cannot be eliminated
because the corresponding momentum expressions are constraints. The
elimination of $\dot\phi$ and $\dot\phi^\dagger$ by using the
expressions (2.3b) and (2.3c) gives

$$\eqalignno{{\cal H}_c=\pi^0\dot A_0
+\bigl(\pi^i-{\kappa\over4\pi}\,
\epsilon^{ij}\,A_j\bigr)\,\dot A_i
&+\bigl(i\pi^\dagger\phi^\dagger-i\pi\phi
-{\kappa\over2\pi}\,\epsilon^{ij}\partial_iA_j\bigr)\,A_0\cr
&+\pi^\dagger\pi-(D_i\phi)^\dagger(D^i\phi)\,.&(2.6)\cr}$$

\bigskip
\noindent Going back with this result to (2.4) and observing the constraint
expressions (2.2) we notice that $\dot A_0$, $\dot A_i$ and $A_0$ can
be respectively absorbed by the Lagrange multipliers $\lambda_0$,
$\lambda_i$ and $\lambda$. After that, the component $A_0$
disappears. This means that it is effectively like we do not have
this degree of freedom any more. In this way it does not make sense
to keep the constraint $\chi^0$ in the theory. Let us the disregard
it (taking $\lambda_0=0$) and consider the total Hamiltonian just as

$${\cal H}=\lambda_i\chi^i+\lambda\psi
+\pi^\dagger\pi-(D_i\phi)^\dagger(D^i\phi)\,.\eqno(2.7)$$

\bigskip
The remaining constraints $\chi^i$ and $\psi$ are apparently
second-class. However, this is not so because the constraint matrix
formed with $\psi$, $\psi^i$ is obvious singular. Actually, a proper
linear combination of them is first-class:

$$\chi=\partial_i\pi^i
+i\bigl(\pi\phi-\pi^\dagger\phi^\dagger\bigr)
+{\kappa\over4\pi}\,\epsilon^{ij}\partial_iA_j\,.
\eqno(2.8)$$

\bigskip
\noindent From now on the set of constraints we shall
use is formed by $\chi$ (first-class) and $\chi^i$ (second-class).

\medskip
At this stage, quantization can be carried out along some lines.  For
instance, gauge freedom associated to first-class constraints can be
frozen by choosing proper gauge-fixing conditions and calculating the
Dirac brackets in order to get commutators~[4,~5]. Another procedure
is to keep the gauge freedom until the final stage of the
quantization~[3,~5].  In this scenario, the physical space of states
have to be extracted by imposing that physical states are annihilated
by first-class constraints, written as operators, while second-class
constraints lead to operational identities. We can also mention the
procedure due to Batalin and Fradkin (BF)~[8] where second class
constraints are transformed into first class by introducing extra
degrees of freedom in the theory~[9].

\vskip1cm
{\titlea III. Effective Lagrangian from the BFV formalism}
\bigskip
The procedure which seems to be relevant for our purposes is the BFV
Hamiltonian path integral formalism, where the gauge choice can be
controlled at any stage due to the powerful Fradkin-Vilkovisky (FV)
theorem~[6]. It assures that the functional generator is independent
of the gauge-fixing since some conditions are satisfied.  Our final
goal is to show that Poincar\'e invariance can be achieved in any
gauge, consistent with the FV conditions.

\medskip
The implementation of the BFV formalism requires the existence of
just first-class constraints. In the present case, this can be
achieved in two ways. The first one is making use of the BF formalism
and transforming second-class constraints into first-class ones.
Another procedure, that we shall develop here, is to eliminate the
second-class constraints with the use of the Dirac bracket definition
and at the same time modifying the path integral measure by means of
the Senjanovic procedure~[10]. This means that the closure of the
algebra of first-class constraints is achieved by using (partial)
Dirac brackets instead of Poisson ones. Let us then write the
constraint matrix element $C^{ij}$

$$\eqalignno{C^{ij}(\vec x,\,\vec y\,)
&=\bigl\{\chi^i(\vec x,t),\,\chi^j(\vec y,t)\bigr\}\,,\cr
&=-{\kappa\over2\pi}\,\epsilon^{ij}\,
\delta^{(2)}(\vec x-\vec y\,)\,.&(3.1)\cr}$$

\bigskip
\noindent Hence,

$$C_{ij}^{-1}(\vec x,\vec y\,)={2\pi\over\kappa}\,\epsilon_{ij}\,
\delta^{(2)}(\vec x-\vec y\,)\,.\eqno(3.2)$$

\bigskip
\noindent Next, using the Dirac bracket definition, we calculate the
fundamental star brackets.

$$\eqalignno{\{A^i(\vec x,t),\, A^j(\vec y,t)\}^\ast
&={2\pi\over\kappa}\,\epsilon^{ij}\,
\delta^{(2)}(\vec x-\vec y\,)\,,&(3.3a)\cr
\{\pi^i(\vec x,t),\, \pi^j(\vec y,t)\}^\ast
&={\kappa\over8\pi}\,\epsilon^{ij}\,
\delta^{(2)}(\vec x-\vec y\,)\,,&(3.3b)\cr
\{A^i(\vec x,t),\, \pi^j(\vec y,t)\}^\ast
&={1\over2}\,\eta^{ij}\,
\delta^{(2)}(\vec x-\vec y\,)\,.&(3.3c)\cr}$$

\bigskip
\noindent Remaining brackets are the same as Poisson ones.

\medskip
Since constraints $\chi^i$ were eliminated, the only remaining
constraint is $\chi$. The algebra satisfied by this constraint is
obtained by using the fundamental Dirac brackets above. We easily see
that it is a rank zero algebra in a sense that

$$\bigl\{\chi(\vec x,t),\,\chi(\vec y,t)\bigr\}^\ast=0\,.\eqno(3.4)$$

\bigskip
\noindent Of course, the expression for the constraint $\chi$ can now
be written by eliminating $\pi_i$ with the use of $\chi_i$, i.e.

$$\chi=i\bigl(\pi\phi-\pi^\dagger\phi^\dagger\bigr)
+{\kappa\over2\pi}\,\epsilon^{ij}\,\partial_iA_j\,.
\eqno(3.5)$$

\bigskip
In terms of the BFV treatment, extra fields have to be incorporated
into the formalism. These are the canonical momentum $p$ of the
Lagrange multiplier $\lambda$, a pair of ghosts $(c,\bar c)$ (related
to the first-class constraint $\chi$) and their corresponding momenta
$(\bar{\cal P},{\cal P})$. The expression of the BRST charge~[6]
reads

$$\Lambda=\int d^2\vec x\,
\bigl(c\chi-i{\cal P}p\bigr)\,.\eqno(3.6)$$

\bigskip
\noindent The transformations of fields generated by this charge in
terms of the star brackets (3.3) are

$$\eqalignno{\delta A_i(x)&=
\bigl\{A_i(\vec x,t),\epsilon\Lambda\bigr\}^\ast\,,\cr
&=-\epsilon\,\partial_i\,c\,(x)\,,\cr
\delta\lambda(x)&=-i\epsilon\,{\cal P}(x)\,,\cr
\delta p\,(x)&=0\,,\cr
\delta\phi(x)&=i\epsilon\,\phi(x)\,c\,(x)\,,\cr
\delta\phi^\dagger(x)&=-i\epsilon\,\phi^\dagger(x)\,c\,(x)\,,\cr
\delta\pi(x)&=-i\epsilon\,\pi(x)\,c\,(x)\,,\cr
\delta\pi^\dagger(x)&=i\epsilon\,\pi^\dagger(x)\,c\,(x)\,,\cr
\delta c\,(x)&=0\,,\cr
\delta\bar c\,(x)&=-i\epsilon\,p\,(x)\,.&(3.7)\cr}$$

\bigskip
Let us now discuss the elimination of the second-class constraints
$\chi^i$ in the generating functional of the BFV formalism.
Considering that~[6]

$$Z=N\int[d\mu^\prime]\,
e^{iS^\prime_{\rm eff}}\,,\eqno(3.8)$$

\bigskip
\noindent where the effective action reads

$$\eqalignno{S^\prime_{\rm eff}=\int d^3x\,\Bigl[\dot A^i\pi_i
&+\dot\phi\,\pi+\dot\phi^\dagger\pi^\dagger+\dot\lambda\,p
+\dot c\,\bar{\cal P}+\dot{\bar c}\,{\cal P}\cr
&+(D_i\phi)^\dagger D^i\phi-\pi^\dagger\pi
+\bigl\{\Psi,\Lambda\bigr\}^\ast\Bigr]\,.&(3.9)\cr}$$

\bigskip
\noindent Here, $\Psi$ is the gauge-fixing function. The relative
position of velocities in the expression above is because we are
using left derivatives for fermionic fields.

\medskip
The elimination of the second-class constraints is achieved by means
of the Senjanovic procedure. This is embodied with the use of the
following measure~[10]

$$\eqalignno{[d\mu^\prime]=\bigl({\rm det}\vert\{\chi_i,
\chi_j\}\vert\bigr)^{1/2}\,\prod_i\,\delta[\chi_i]
[dA_i]&[d\pi_i][d\phi][d\phi^\dagger][d\pi][d\pi^\dagger]\cr
\times&[d\lambda][dp][dc][d\bar c]
[d{\cal P}][d\bar{\cal P}]\,.&(3.10)\cr}$$

\bigskip
\noindent The determinant factor in the expression above does not
involve any field. It is then possible to absorb it into the
normalization factor $N$. We also use the functional integration over
$\pi_i$ to eliminate the second-class constraint. The result is a
simpler expression for the vacuum functional

$$Z=N\int[d\mu]\,e^{iS_{\rm eff}}\,,\eqno(3.11)$$

\bigskip
\noindent where

$$\eqalignno{[d\mu]&=[dA_i][d\phi][d\phi^\dagger][d\pi]
[d\pi^\dagger][d\lambda][dp][dc][d\bar c][d{\cal P}]\,,&(3.12)\cr
S_{\rm eff}&=\int d^3x\,\Bigl[{\kappa\over4\pi}\,
\epsilon^{ij}\dot A_iA_j
+\dot\phi\,\pi+\dot\phi^\dagger\pi^\dagger+\dot\lambda\,p
+\dot c\,\bar{\cal P}+\dot{\bar c}\,{\cal P}\cr
&\phantom{=\int d^3x\,\Bigl[{\kappa\over4\pi}\,
\epsilon^{ij}\dot A_iA_j}
+(D_i\phi)^\dagger D^i\phi
-\pi^\dagger\pi+\bigl\{\Psi,\Lambda\bigr\}^\ast
\Bigr]\,.&(3.13)\cr}$$

\bigskip
\noindent The effective action is BRST invariant independently of the
choice of $\Psi$. This can be easily verified by using the Jacobi
identity and the nilpotency of the BRST charge.

\medskip
In order to see if the procedure we have used above to eliminate the
second-class constraints is consistent or not we could choose some
particular gauge, develop the integration of the vacuum functional
and compare the result with the ones found in literature. We show
this in the Appendix A, where some comments are also made in order to
become clearer what will be developed in the next sections.

\vskip1cm
{\titlea IV. Poincar\'e invariance and absence of anomalies}
\bigskip
As it was seen in the Appendix A, to integrate over the momenta it is
necessary to choose a particular gauge-fixing condition. Choosing
$\Psi=i\bar c\,\bigl({1\over2}\alpha p -\nabla\!\cdot\!\vec A\bigr)
+\bar{\cal P}\lambda$ and integrate over $\pi$, $\pi^\dagger$, $\cal
P$, $\bar{\cal P}$ we obtain the effective Lagrangian used by Shim et
al.~[2] where it was detected anomaly in the angular momentum
algebra. In order to make an analysis in a more general point of view,
let us consider here the initial effective Lagrangian (3.13), where
no particular gauge-fixing function has yet been chosen. Now, the
variables $p$, $\pi$, $\pi^\dagger$, $\cal P$ and ${\cal P}$ do not
have to be seen as momenta, but just as variables of an extended
configuration space. If $\{\Psi,\Lambda\}^\ast$ does not contain
velocities (this is not the case, for example, of the Fock-Schwinger
gauge - see Appendix A), we can identify the usual bracket relations
involving these variables in terms of the Dirac brackets. For
instance, considering that $\{\Psi,\Lambda\}^\ast$ does not contain
$\dot\lambda$ we have $\partial{\cal L}/\partial\dot\lambda=p$ and
$\partial{\cal L}/\partial\dot p=0$. Both are second class
constraints. Using the Dirac brackets definition we get
$\{\lambda(\vec x,t),p(\vec y,t)\}^\ast=\delta^{(2)}(\vec x-\vec
y\,)$.  If $\{\Psi,\Lambda\}$ had dependence on $\dot\lambda$, the
effective momentum conjugate to $\lambda$ would be $\partial{\cal
L}/\partial\dot\lambda=p+\partial\{\Psi,\Lambda\}^\ast/\partial\dot\lambda$.
This is an important fact as we are going to see soon. The absence of
anomaly in the Poincar\'e algebra, for any $\Psi$, will be related to
it. If one changes the gauge, momenta may also change in order to
preserve the symmetry.

\medskip
Let us study this point with details. Denoting all fields
that appear in (3.13) generically by $\xi_A$ we have that any
on-shell variation of these fields leads to (using left-derivatives)

$$\delta{\cal L}=\partial_\mu\,\Bigl(\delta\xi_A\,
{\partial{\cal L}\over\partial\,
(\partial_\mu\xi_A)}\Bigr)\,.\eqno(4.1)$$

\bigskip
We first discuss invariance under spacetime translations.
Considering infinitesimal translations given by $\delta
x^\mu=x^{\prime\mu}-x^\mu=\epsilon^\mu$ and that, for any field,
$\xi^\prime_A(x^\prime)=\xi_A(x)$ we have

$$\eqalignno{\delta\xi_A(x)&=\xi^\prime_A(x)-\xi_A(x)\,,\cr
&=\xi^\prime_A(x)-\xi_A^\prime(x^\prime)\,,\cr
&=-\,\epsilon^\mu\partial_\mu\,\xi_A(x)
+0(\epsilon^2)\,.&(4.2)\cr}$$

\bigskip
\noindent Of course we have a similar relation for $\delta{\cal L}$.
Introducing these results into (4.1) one obtains the usual expression
for the energy-momentum tensor

$${\cal T}_{\mu\nu}=\partial_\nu\,\xi_A\,
{\partial{\cal L}\over\partial\,(\partial^\mu\xi_A)}
-\eta_{\mu\nu}\,{\cal L}\,,\eqno(4.3)$$

\bigskip
\noindent which is divergenceless with respect the index $\mu$ (in
general, the energy-momentum tensor is not symmetric). Let us then
check if

$$P_\nu=\int d^2x\,{\cal T}_{0\nu}\eqno(4.4)$$

\bigskip
\noindent is actually the generator of spacetime translations.
Considering first $P_0$ we have

$$\eqalignno{\delta\xi_A(x)
&=\epsilon^0\bigl\{P_0,\,\xi_A(x)\bigr\}^\ast\,,\cr
&=\epsilon^0\int d^2\vec y\,\Bigl(
\dot\xi_B(\vec y,t)\bigl\{
{\partial{\cal L}\over\partial\dot\xi_B(\vec y,t)},\,
\xi(\vec x,t)\bigr\}^\ast\cr
&\phantom{=\epsilon^0\int d^2\vec y\,}
+\bigl\{\dot\xi_B(\vec y,t),\,\xi(\vec x,t)\bigr\}^\ast\,
{\partial{\cal L}\over\partial\dot\xi_B(\vec y,t)}\cr
&\phantom{=\epsilon^0\int d^2\vec y\,}
-\bigl\{{\cal L}(\vec y,t),\,\xi(\vec x,t)\bigr\}^\ast
\Bigr)\,,\cr
&=-\,\epsilon^0\dot\xi(x)\,.&(4.5)\cr}$$

\bigskip
\noindent In the last step above we have used the identity

$$\{{\cal L}(\vec y,t),\,\xi(\vec x,t)\}^\ast
={\partial{\cal L}\over\partial\dot\xi_B(\vec y,t))}\,
\{\dot\xi_B(\vec y,t),\,\xi(\vec x,t)\}^\ast\,.\eqno(4.6)$$

\medskip
\noindent For pure space translations we have

$$\eqalignno{\delta\xi_A(x)
&=\epsilon^i\bigl\{P_i,\,\xi_A(x)\bigr\}^\ast\,,\cr
&=\epsilon^i\int d^2\vec y\,
\bigl\{\partial_i\xi_B(\vec y,t)\,
{\partial{\cal L}\over\partial\dot\xi_B(\vec y,t)},\,
\xi_A(\vec x,t)\bigr\}^\ast\,,\cr
&=-\epsilon^i\partial_i\,\xi_A(x)\,.&(4.7)\cr}$$

\bigskip
\noindent Since there are no problems with ordering operators, all
the (Dirac) brackets above can be replaced by commutators and the
corresponding relations can be seen quantically. There is no anomaly.
It might be opportune to mention that there was no anomaly for
spacetime translations in the previous mentioned paper~[2].

\medskip
Let us now discuss the invariance under spacetime rotations. This
cannot be obtained so directly as the translation case. It is
necessary to know the Lorentz transformations of fields which appear
in (3.13). Since it is written in a manifestly noncovariant form, we
have to figure out what must be the their Lorentz transformations in
order to preserve $\cal L$ as a scalar (see Appendix B). The problem
here is that there is a lot of fields that we do not know their
Lorentz transformation and it is a very difficult task to figure them
out.  Fortunately, we do not need to know the transformation of all
them, only those  with spacetime derivatives (see expression (4.1)).
Since $p$, $\pi$, $\pi^\dagger$, $\cal P$ and $\bar{\cal P}$ are
auxiliary fields (not dynamical), we just need to know the
transformations of $A_i$, $\lambda$, $\phi$, $\phi^\dagger$, $c$ and
$\bar c$.

\medskip
The transformation of scalar fields reads (see Appendix B)

$$\eqalignno{\delta\phi(x)
&=-\,\omega^{\mu\nu}x_\nu\partial_\mu\,\phi(x)\,,\cr
\delta\phi^\dagger(x)
&=-\,\omega^{\mu\nu}x_\nu\partial_\mu\,\phi(x)^\dagger\,.&(4.8)\cr}$$

\bigskip
\noindent Concerning the gauge field we have

$$\eqalignno{\delta A_i(x)&=A^\prime_i(x)-A_i(x)\,,\cr
&=-\,\omega^{\mu\nu}x_\nu\partial_\mu A_i(x)
+\omega_i^\mu A_\mu(x)\,.&(4.9)\cr}$$

\bigskip
\noindent As one observes, the transformation of $A_i$ requires the
existence of $A_0$, that was disregarded in the beginning. However,
this is not a problem because in the BFV formalism one always
make $A_0$ appears again by identifying $\lambda$ as $A_0$,
independently of the gauge choice (see Appendix A). So, from now on
we replace $\lambda$ by $A_0$.

\medskip
Of course, $D_i\phi$ and $(D_i\phi)^\dagger$ have the same
transformation as $A_i$. Namely,

$$\eqalignno{\delta\,\bigl(D_i\phi(x)\bigr)
&=-\,\omega^{\mu\nu}x_\nu\partial_\mu\bigl(D_i\phi(x)\bigr)
+\omega_i^\mu D_\mu\phi(x)\,,\cr
\delta\,\bigl(D_i\phi(x)\bigr)^\dagger
&=-\,\omega^{\mu\nu}x_\nu\partial_\mu\bigl(D_i\phi(x)\bigr)^\dagger
+\omega_i^\mu(D_\mu\phi(x))^\dagger\,.&(4.10)\cr}$$

\bigskip
Concerning the transformation of $c$ and $\bar c$, we report back to
the particular cases described in the Appendix A. With those
particular gauge choices, we see that ghost behave as scalars. Since
this intrinsic property cannot depend on gauge choices we conclude
that $c$ and $\bar c$ must always transform as scalars
\footnote{(*)}{The same argument could not be applied to the
auxiliary fields $p$, $\pi$, $\pi^\dagger$, $\cal P$ and $\bar{\cal
P}$. For example, for the first gauge choice seen in the Appendix A,
$p$ transforms as a scalar. For the second one, where
$\{\Psi,\Lambda\}$ contains $\dot\lambda$, $p$ does not transform as
scalar anymore.}, so

$$\eqalignno{\delta\,c\,(x)
&=-\,\omega^{\mu\nu}x_\nu\partial_\mu c\,(x)\,,\cr
\delta\,\bar c\,(x)
&=-\,\omega^{\mu\nu}x_\nu\partial_\mu\bar c\,(x)\,.&(4.11)\cr}$$

\bigskip
\noindent Considering the Lagrangian (3.13) we obtain that expression
(4.1) leads to

$$\delta{\cal L}=\partial_\mu\Bigl[
\delta A_\nu\,
{\partial{\cal L}\over\partial\,(\partial_\mu A_\nu)}
+\delta\phi\,
{\partial{\cal L}\over\partial\,(\partial_\mu\phi)}
+\delta\phi^\dagger\,
{\partial{\cal L}\over\partial\,(\partial_\mu\phi^\dagger)}
+\delta c\,
{\partial{\cal L}\over\partial\,(\partial_\mu c)}
+\delta\bar c\,
{\partial{\cal L}\over\partial\,(\partial_\mu\bar c)}\Bigr]\,.
\eqno(4.12)$$

\bigskip
\noindent Replacing the transformations (4.8)-(4.10) into the expression
above, one obtains the following divergenceless quantity (in the
$\mu$ index)

$$\eqalignno{{\cal M}_{\mu\nu\rho}
=\bigl(x_\lambda\partial_\rho-x_\rho&\partial_\lambda\bigr)
\Bigl(A_\nu\,{\partial{\cal L}\over\partial\,(\partial^\mu A_\nu)}
+\phi\,{\partial{\cal L}\over\partial\,(\partial_\mu\phi)}
+\phi^\dagger\,{\partial{\cal L}\over\partial\,
(\partial_\mu\phi^\dagger)}
+c\,{\partial{\cal L}\over\partial\,(\partial_\mu c)}
+\bar c\,{\partial{\cal L}\over\partial\,
(\partial_\mu\bar c)}\Bigr)\cr
&\,+\Bigl(\eta_{\lambda\nu}A_\rho+\eta_{\rho\nu}A_\lambda\Bigr)\,
{\partial{\cal L}\over\partial\,(\partial^\mu A_\nu)}
+\bigl(\eta_{\mu\lambda}x_\rho
-\eta_{\mu\rho}x_\lambda\bigr)\,{\cal L}\,.&(4.13)\cr}$$

\bigskip
\noindent We are interested in the quantities

$$M_{\lambda\rho}=\int d^2x\,{\cal M}_{0\lambda\rho}\,,\eqno(4.14)$$

\bigskip
\noindent which have to be the generators of the Lorentz algebra.
{}From (4.13) we get

$$\eqalignno{M_{\lambda\rho}=\int d^2\vec x\,\Bigl[
\bigl(x_\lambda\partial_\rho-x_\rho\partial_\lambda\bigr)\,
\Bigl(A_\nu\,{\partial{\cal L}\over\partial\dot A_\nu}
+\phi\,{\partial{\cal L}\over\partial\dot\phi}
&+\phi^\dagger\,{\partial{\cal L}\over\partial\dot\phi^\dagger}
+c\,{\partial{\cal L}\over\partial\dot c}
+\bar c\,{\partial{\cal L}\over\partial\dot{\bar c}}\Bigr)\cr
&+\bigl(\eta_{\lambda\nu}A_\rho
-\eta_{\rho\nu}A_\lambda\bigr)\,
{\partial{\cal L}\over\partial\dot A_\nu}\cr
&+\bigl(\eta_{\lambda 0}x_\rho-\eta_{\rho 0}x_\lambda\bigr)\,
{\cal L}\Bigr]\,.&(4.15)\cr}$$

\bigskip
The quantities $\partial{\cal L}/\partial\dot A_\nu$, $\partial{\cal
L}/\partial\dot \phi$ etc. are the conjugate momenta related to
$A_\nu$, $\phi$ etc.. They are not necessarily the same as $\pi^\nu$,
$\pi$ etc. So, the usual Lorentz algebra involving $M_{\lambda\rho}$
can be verified trivially. Since there are no problem with ordering
operators, the same algebra can be directly written in terms of
commutators. There is no anomaly for any gauge we choice. This result
is in agreement with the one recently found by Banerjee~[3].

\vskip 1cm
{\titlea V. Conclusion}
\bigskip
We have studied the Abelian CS theory coupled to matter field. Our
starting point was the initial effective Lagrangian of the BFV
formalism. The importance of this Lagrangian is because the
gauge-fixing function has not yet been chosen and we know that the
generating functional is independent of this choice if some
conditions are satisfied (Fradkin-Vilkovisky theorem). So, the
results we have obtained are independent of any gauge choice we make.
We have then show that the closure of the full Poincar\'e algebra is
achieved without any anomaly, in agreement with a recent paper of
Banerjee~[3] and contrarily to previous publications~[2]. To show
this we have made a careful analysis of the Noether currents and the
momenta associated to the dynamical variables of the theory.

\vskip 1cm
{\titlea Acknowledgment}
\bigskip
This work was supported in part by Conselho Nacional de
Desenvolvimento Cient\\fi- co e Tecnol\'ogico - CNPq (Brazilian
Research Agency).

\vskip 1cm
{\titlea Appendix A}
\bigskip
In order to verify the consistency of what we have done in Sec. III,
let us first make an usual choice for the gauge-fixing function
$\Psi$, that is to say~[6]

$$\Psi=i\bar c\,\bigl({1\over2}\alpha p
-\nabla\!\cdot\!\vec A\,\bigr)
+\bar{\cal P}\lambda\eqno(A.1)$$

\bigskip
\noindent and develop expression (3.11) by integrating over the
momenta. We shall obtain a covariant effective action with the
Lorentz-like term $(\partial_\mu A^\mu)^2/2\alpha$. First we notice
that the gauge choice given by (A.1) leads to

$$\bigl\{\Psi(x),\,\Lambda\bigr\}^\ast
=-{\alpha\over2}\,p^2-i\bar c\,\nabla^2c
+p\,\nabla\!\cdot\!\vec A-i\bar{\cal P}{\cal P}
-\lambda\,\Bigl[i\bigl(\pi\phi
-\pi^\dagger\phi^\dagger\bigr)+{\kappa\over2\pi}\,
\epsilon^{ij}\partial_iA_j\Bigr]\,.\eqno(A.2)$$

\bigskip
\noindent With this result the effective action turns to be

$$\eqalignno{S_{\rm eff}=\int d^3x\,\Bigl\{
{\kappa\over4\pi}\,\epsilon^{ij}\dot A_iA_j
&+\dot\phi\,\pi+\dot\phi^\dagger\pi^\dagger
+\dot\lambda\,p+\dot c\,\bar{\cal P}
+\dot{\bar c}\,{\cal P}
+(D_i\phi)^\dagger D^i\phi\cr
&-\pi^\dagger\pi-{\alpha\over2}p^2-i\bar c\,\nabla^2c
+p\,\nabla\!\cdot\!\vec A
-i\bar{\cal P}{\cal P}\cr
&-\lambda\,\Bigl[i\bigl(\pi\phi
-\pi^\dagger\phi^\dagger\bigr)+{\kappa\over2\pi}\,
\epsilon^{ij}\partial_iA_j\Bigr]\Bigr\}\,.&(A.3)\cr}$$

\bigskip
\noindent Going back to the path integral and integrating over the
momenta we obtain

$$\eqalignno{Z=N\int[dA_i]&[d\phi][d\phi^\dagger]
[d\lambda][dc][d\bar c]\,\exp\,\Bigl\{i\int d^3x
\Bigl[{\kappa\over4\pi}\,\epsilon^{ij}\dot A_iA_j
+(D_i\phi)^\dagger D^i\phi
-i\bar c\,\sq c\cr
&+(\dot\phi+i\lambda)(\dot\phi^\dagger-i\lambda)
+{1\over2\alpha}\,(\dot\lambda+\nabla\!\cdot\!\vec A)^2
-{\kappa\over2\pi}\,\lambda\,
\epsilon^{ij}\partial_iA_j\Bigr]\Bigr\}\,.&(A.4)\cr}$$

\bigskip
\noindent In the BFV formalism the manifest covariance is obtained by
always letting $\lambda$ to be the missing temporal component of the
gauge field. Doing this we finally get

$$\eqalignno{Z=N\int[dA_\mu][d\phi][d\phi^\dagger]
&[dc][d\bar c]\,
\exp\,\Bigl\{i\int d^3x\,\Bigl[{\kappa\over4\pi}\,
\epsilon^{\mu\nu\rho}\,A_\mu\partial_\nu A_\rho\cr
&+(D_\mu\phi)^\dagger D^\mu\phi+i\bar c\,\sq c
+{1\over2\alpha}\,(\partial_\mu A^\mu)^2\Bigr]\Bigr\}\,.
&(A.5)\cr}$$

\bigskip
\noindent If we had not integrated on the momentum $p$ we would have
obtained

$$\eqalignno{Z=N\int[dA_\mu][d\phi][d\phi^\dagger]
&[dc][d\bar c][dp]\,
\exp\,\Bigl\{i\int d^3x\,\Bigl[{\kappa\over4\pi}\,
\epsilon^{\mu\nu\rho}\,A_\mu\partial_\nu A_\rho\cr
&+(D_\mu\phi)^\dagger D^\mu\phi+i\bar c\,\sq c
-{\alpha\over2}\,p^2+p\,\partial_\mu A^\mu\Bigr]\Bigr\}\,.
&(A.6)\cr}$$

\bigskip
\noindent This is precisely the Lagrangian used as the starting point
in the paper by H. Shin et al.~[2] where it was detected an
anomaly in the Poincar\'e algebra.

\bigskip
Although not common, it could have chosen another gauge-fixing
function different from (A.1). With the purpose of illustrating a
little more this subject and for future reference, let us consider
another gauge-fixing function. We choose the one based on the
Fock-Schwinger gauge~[12], $x_\mu A^\mu=0$.  To implement this gauge
in the BFV formalism we have to choose $\Psi$ as~[13]

$$\Psi=i\bar c\,\dot\lambda-i\dot{\bar c}\,\lambda
+i\bar c\,\Bigl({1\over2}\alpha p
+{x_\mu A^\mu\over x^2}\Bigr)
+\bar{\cal P}\lambda\eqno(A.7)$$

\bigskip
\noindent and consider $\lambda$ such that $\{\lambda,\dot
p\}^\ast=0$.  Note here that $\Psi$ has dependence on the velocity
$\dot\lambda$.  Integrating over the momenta and identifying
$\lambda$ as $A_0$ we get

$$\eqalignno{Z=N\int[dA_\mu][d\phi][d\phi^\dagger]
&[dc][d\bar c]\,
\exp\Bigl\{i\int d^3x\,\Bigl[{\kappa\over4\pi}\,
\epsilon^{\mu\nu\rho}\,A_\mu\partial_\nu A_\rho
+(D_\mu\phi)^\dagger D^\mu\phi\cr
&+i\bar c\,{x^\mu\partial_\mu\over x^2}\,c
+{1\over2\alpha x^4}\,(x_\mu A^\mu)^2\Bigr]\Bigr\}\,.
&(A.8)\cr}$$

\vskip 1cm
{\titlea Appendix B}
\bigskip
In order to see how one can figure out the Lorentz transformation of
fields in manifestly noncovariant Lagrangians, let us consider a
simple example just involving real scalar fields

$${\cal L}=\pi\dot\phi+{1\over2}\,
\partial_i\phi\,\partial^i\phi
-{1\over2}\,\pi^2\,.\eqno(B.1)$$

\bigskip
\noindent We already know the Lorentz transformation of $\phi$. Since
it is a scalar, we have

$$\phi^\prime(x^\prime)=\phi(x)\,.\eqno(B.2)$$

\bigskip
\noindent For an infinitesimal Lorentz transformation $\delta
x^\mu=\omega^{\mu\nu}x_\nu$ and using (B.2) we obtain

$$\eqalignno{\delta\phi(x)&=\phi^\prime(x)-\phi(x)\,,\cr
&=\phi^\prime(x)-\phi^\prime(x^\prime)\,,\cr
&=-\,\omega^{\mu\nu}x_\nu\partial_\mu\phi(x)
+0(\omega^2)\,.&(B.3)\cr}$$

\bigskip
\noindent This is the transformation that characterizes scalar
fields. Since $\cal L$ is also supposed to be a scalar it must
transform in the same way. So what we have to do is to figure out a
transformation for $\pi$ in order that this occurs. Considering (B.1)
and (B.3) we have

$$\eqalignno{\delta{\cal L}&=\delta\pi\dot\phi
+\pi\delta\dot\phi+\partial_i\phi\,\partial^i\delta\phi
-\pi\delta\pi\,,\cr
&=-\,\omega^{\mu\nu}x_\nu\partial_\mu\,\bigl(\pi\dot\phi
+{1\over2}\,\partial_i\phi\,\partial^i\phi\bigr)
+\bigl(\dot\phi-\pi)\,\delta\pi\cr
&\phantom{=}+\omega^{\mu\nu}x_\nu\partial_\mu\pi\dot\phi
+\omega^{i0}\partial_i\phi\,\bigl(\dot\phi-\pi\bigr)\,.&(B.4)\cr}$$

\bigskip
\noindent Observing the expression above, one concludes that the
transformation for $\pi$ that renders $\cal L$ a scalar
transformation is

$$\delta\pi=-\,\omega^{\mu\nu}x_\nu\partial_\mu\pi
-\omega^{0i}\partial_i\phi\,.\eqno(B.5)$$

\bigskip
\noindent We notice that it is the same transformation as $\dot\phi$
as it must be.

\vfill\eject
{\titlea References}
\bigskip
\item {[1]} For a recent review on CS theories, see S. Forte, Rev.
Mod. Phys. 64 (1992) 193 and references therein.
\item {[2]} See H. Shin, W.-T. Kim. J.-K. Kim and Y.-J. Park, Phys.
Rev. D46 (1992) 2730 and references therein.
\item {[3]} R. Banerjee, Phys. Rev. D48 (1993) 2905.
\item {[4]} P.A.M. Dirac, Can. J. Math. 2 (1950) 129; {\it Lectures
on quantum mechanics} (Yeshiva University, New York, 1964).
\item {[5]} See, for example, A. Hanson, T. Regge and C. Teitelboim:
{\it Constrained Hamiltonian systems} (Academia Nazionale dei Lincei,
Rome, 1976).
\item {[6]} E.S. Fradkin and G.A. Vilkovisky, Phys. Lett. B55 (1975)
224; I.A. Batalin and G. Vilkovisky, Phys. Lett. B69 (1977) 309; For
a review see M. Henneaux, Phys. Report 126 (1985) 1.
\item {[7]} G.W. Semmenoff and P. Sodano, Nucl. Phys. B328 (1989) 753;
R. Banerjee, A. Chatterjee, and V.V. Sreedhar, Ann. Phys. (N.Y.) 222
(1993) 254.
\item {[8]} L.D. Faddeev and S.L. Shatashvili, Phys. Lett. B167
(1986) 225; I.A. Batalin and E.S. Fradkin, Phys. Lett. B180 (1986)
157; Nucl. Phys. B279 (1987) 514; Nucl. Phys. B326 (1989) 701. See
also, I.A. Batalin and I.V. Tyutin, Int. Jour. Mod. Phys. A6 (1991)
3255 and other references therein.
\item {[9]} R. Banerjee, Phys. Rev. D48 (1993) R5467.
\item {[10]} A.P. Senjanovic, Ann. Phys. (N.Y.) 100 (1976) 227.
\item {[11]} C. Becchi, A. Rouet and R. Stora, Ann. Phys. (NY) 98
(1976) 287; I.V. Tyutin, Lebedev preprint FIAN-39/1975, unpublished;
\item {[12]} V.A. Fock, Sov. Phys. 12 (1937) 404; J. Schwinger, Phys.
Rev. 82 (1952) 684. See also W. Kummer and J. Weiser, Z. Phys. 31C
(1986) 105.
\item {[13]} J. Barcelos-Neto, C.A.P. Galv\ao and P. Gaete, Mod.
Phys. Lett. A6 (1991) 1597.

\bye